\documentclass{ws-procs975x65}
\begin{document}

\title{POVMs: a small but important step\\beyond standard quantum mechanics}

\author{Willem M. de Muynck}

\address{Theoretical Physics,
Eindhoven University of Technology\\
Eindhoven, the Netherlands,\\
E-mail: w.m.d.muijnck@tue.nl\\
http://www.phys.tue.nl/ktn/Wim/muynck.htm}

\begin{abstract}
It is the purpose of the present contribution to demonstrate that
the generalization of the concept of a quantum mechanical
observable from the Hermitian operator of standard quantum
mechanics to a positive operator-valued measure is not a
peripheral issue, allegedly to be understood in terms of a trivial
nonideality of practical measurement procedures, but that this
generalization touches the very core of quantum mechanics, viz.
complementarity and violation of the Bell inequalities.
\end{abstract}


\bodymatter

\section{Introduction}\label{sec1}
I shall refer to the usual quantum mechanical formalism dealt with
in quantum mechanics textbooks as the \textit{standard} formalism.
In this formalism a quantum mechanical (standard) observable is
represented by a Hermitian operator, having a spectral
representation consisting of projection operators $E_m$, which
constitute an \textit{orthogonal} $(Tr E_mE_{m'} = O,\; m\neq m')$
decomposition of the identity operator,
\[E_{m}\geq O,\; E_m^2 =E_m,\;\sum_{m} E_{m} = I.\]
The projection operators $E_m$ are said to generate a
projection-valued measure (PVM). A \textit{generalized} observable
is represented by a positive operator-valued measure (POVM),
generated by a set of non-negative operators $M_m$ that in general
are \textit{not} projection operators and constitute a
\textit{non}-orthogonal decomposition of the identity operator, i.
e., in general $Tr M_mM_{m'} \neq O$, and
\begin{equation}M_{m}\geq O, \;\sum_{m} M_{m} = I.\label{1}
\end{equation}
In generalized quantum mechanics measurement probabilities are
given according to
\begin{equation} p_m= Tr \rho M_m,\label{2}
\end{equation}
$\rho$ a density operator, and the set $\{M_m\}$ satisfying
(\ref{1}). In the following generalized and standard observables
will be referred to by their POVM and PVM, respectively.

Nowadays it is more and more realized that standard quantum
mechanics is not completely adequate for dealing with quantum
information, and that it is necessary to consider POVMs. Thus, as
is well known, standard observables can only yield information on
\textit{diagonal} elements of $\rho.$ Since in general the
operators $M_m$ of a POVM need not commute with each other, POVMs
may yield information on \textit{off-diagonal} elements of $\rho,$
too. Particularly interesting is the existence of
\textit{complete} POVMs allowing to reconstruct the density
operator from the set of probabilities (\ref{2}) obtained in a
measurement of one single \textit{generalized} observable. By
means of `quantum tomography' state reconstruction can be achieved
also in standard quantum mechanics; however, in general this
method requires measurement of a large number of standard
observables.

POVMs are obtained in a natural way when applying quantum
mechanics to the interaction of object and measuring instrument.
Thus, let $\rho^{(o)}$ and $\rho^{(a)}$ be the initial density
operators of object and measuring instrument, respectively, and
let $\rho^{(oa)}_{fin} = U\rho^{(o)}\otimes \rho^{(a)}
U^\dagger,\;U = e^{-\frac{i}{\hbar}HT}$ be the final state of the
interaction. If in this latter state a measurement is performed of
the pointer observable $\{E^{(a)}_m\}$, then the measurement
probabilities are found according to
\begin{equation}\label{3} p_{m} = Tr_{oa}\;
\rho^{(oa)}_{fin} E^{(a)}_m= Tr_o \rho^{(o)} M_m,\end{equation}
with $M_m$ given by
\begin{equation}\label{4} M_m =Tr_{a} \rho^{(a)} U^\dagger
E^{(a)}_m U.\end{equation}
 From expression (\ref{4}) it follows that there is no reason to expect that
$M_{m}$
 should be a projection operator; and in general it isn't.

The generalization of the mathematical formalism by means of the
introduction of POVMs entails a considerable extension of the
domain of application of quantum mechanics. In particular, it is
possible that the set of operators $\{M_m\}$ of a POVM spans the
whole of Hilbert-Schmidt space, in which case we have a
\textit{complete} measurement. Such \textit{complete} measurements
are experimentally feasible\cite{dM98,dM2002,AlBaNi04}, and may
have considerable practical importance because of their richer
informational content.

As can be seen from (\ref{3}) and  (\ref{4}), it is the
interaction of object and measuring instrument which is at the
basis of the notion of a POVM. In quantum mechanics textbooks
measurement is generally treated in an axiomatic way, and a
detailed description of it is virtually absent. Bohr was one of
the few to take the problem seriously, but he was mainly
interested in the \textit{macroscopic} phase of the measurement,
which, according to him, was to be described by \textit{classical}
mechanics. However, not the macroscopic but rather the
{\textit{microscopic} phase of the measurement, in which the
microscopic information is transferred from the microscopic object
to the measuring instrument, is crucial for obtaining
\textit{quantum} information. This phase should be described by
quantum mechanics.

The influence of measurement has been of the utmost importance in
the early days of quantum mechanics. In particular has it been a
crucial feature at the inception of the notion of
\textit{complementarity}. In section~\ref{sec2} it will be shown
that POVMs indeed play a crucial role there. By hindsight it can
be concluded that much confusion could have been prevented if at
that time it would have been realized that the standard formalism
of quantum mechanics (as laid down, for instance, in von Neumann's
authoritative book\cite{vN32}) is just a preliminary step towards
a more general formalism. As it is evident now, the standard
formalism is not even able to yield a proper description of the
so-called thought experiments, at that time being at the heart of
our understanding of quantum mechanics\cite{dM2002}.

In this contribution the importance of the generalized formalism
of POVMs is demonstrated by means of two examples, the first one
elucidating Bohr's concept of complementarity in the sense of
mutual disturbance of the measurement results of jointly measured
incompatible standard observables (section~\ref{sec2}). As a
second example, in section~\ref{sec3} a \textit{generalized}
Aspect experiment is discussed using the POVM formalism, thus
demonstrating that violation of the Bell inequalities can be seen
as a consequence of complementarity rather than as being
associated with nonlocality.

\section{Complementarity}\label{sec2}

\subsection{The Summhammer, Rauch, Tuppinger
experiment}\label{sec2.1}

In this section I will now discuss as an example of the
double-slit experiment a neutron interference experiment performed
by Summhammer, Rauch and Tuppinger\cite{SuRaTu}. Other examples
can be found in many different areas of experimental
physics\cite{dM2002}. Let us first consider the two limiting cases
\begin{figure}
\psfig{file=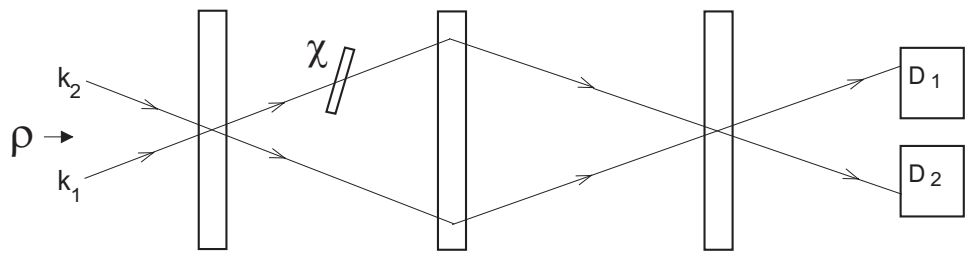,width=5in}
 \caption{Pure interference measurement.}
  \label{fig1}
\end{figure}
which can be treated by means of standard quantum mechanics, viz.
\begin{figure}
\psfig{file=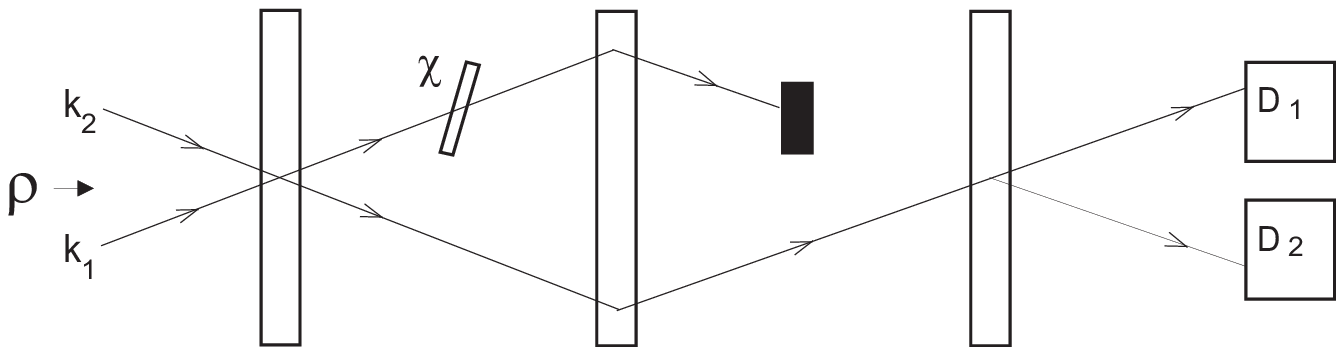,width=5in}
 \caption{Pure path measurement.}
  \label{fig2}
\end{figure}
what I shall denote as the `pure interference measurement' (cf.
figure~\ref{fig1}) and the `pure path measurement' (cf.
figure~\ref{fig2}). In the figures a neutron interferometer is
schematically represented by three vertical slabs in which Bragg
reflection takes place and interference of the different paths can
be realized after a possible phase shift $\chi$ has been applied
in one of the paths. Detectors $\rm D_1$ and $\rm D_2$ can be
placed either behind the third slab (figure~\ref{fig1}), or behind
the second one, in which case the experiment reduces to a
`which-path' measurement (in figure~\ref{fig2} this is realized by
putting an ideal absorber in one path, while adding the
measurement frequencies of detectors $\rm D_1$ and $\rm D_2$ to
obtain the probability $p_+$ the neutron was in the second path;
$p_- = 1-p_+$ is the probability that the neutron was in the path
of the ideal absorber).

The observables measured in the measurement arrangements of
figures~\ref{fig1} and \ref{fig2} are standard observables which
can easily be found on the basis of elementary
considerations\cite{MuMa90}. With $\rho$ the initial (incoming) state we find: \\
\noindent \textit{pure interference measurement}: $p_n = Tr \rho
Q_n,\;n=1,2$, $\{Q_1,Q_2\}$ the PVM of the standard
interference observable;\\
\noindent \textit{pure path measurement}: $p_m = Tr \rho P_m,\;
m=+,-$, $\{P_+,P_-\}$ the PVM of the standard path observable.
\\It will not be necessary to display these observables
explicitly; it is sufficient to know that the operators $Q_n$ and
$P_m$ are projection operators, defining PVMs of standard quantum
mechanics.

In the experiment performed by Summhammer, Rauch and
Tuppinger\cite{SuRaTu} an absorber (transmissivity $a$) is
inserted in one of the paths (figure~\ref{fig3}).
\begin{figure}
\psfig{file=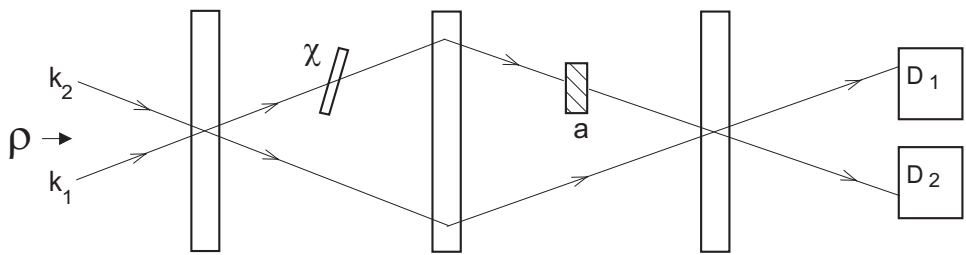,width=5in}
 \caption{Neutron interference measurement by
Summhammer, Rauch and Tuppinger.}
 \label{fig3}
\end{figure}
In the limits $a=1$ and $a=0$ the Summhammer, Rauch, Tuppinger
experiment reduces to the pure interference and the pure path
measurement, respectively. The interesting point is that in
between these limits the experiment is no longer described by a
PVM, but by a POVM. Indeed, we find for $0\leq a\leq 1$: $p_i = Tr
\rho M_i, i=1,2,3,$ in which $i=3$ refers to those events in which
the neutron is absorbed by the absorber. It is easily
seen\cite{MuMa90} that the operators $M_i$ are given according to
\begin{equation}\label{5}\left\{\begin{array}{l}
 M_1=\frac{1}{2}[P_++aP_- + \sqrt{a}(Q_1-Q_2)],\\
 M_2=\frac{1}{2}[P_++aP_- -\sqrt{a}(Q_1-Q_2)],\\
 M_{3}=(1-a)P_-.
 \end{array}\right. \end{equation}

In the following way the measurement represented by the POVM
$\{M_1,M_2,M_3\}$, $M_i$ given by (\ref{5}), can be interpreted as
a joint nonideal measurement of the interference and path
observables defined above. Define the bivariate POVM $(R_{mn})$ as
follows:
\begin{equation} \label{16} (R_{mn}):=\left(\begin{array}{cc}
M_1 & M_2\\
\frac{1}{2}M_{3} & \frac{1}{2}M_{3}
\end{array}\right).\end{equation}
Then the two marginals, $\{\sum_{m} R_{mn}, n=1,2\}$ and
$\{\sum_{n} R_{mn}, m=+,-\},$ are easily found. It directly
follows from (\ref{5}) that these marginals are related to the
PVMs $\{Q_1,Q_2\}$ and $\{P_+,P_-\}$, respectively, according to
\begin{equation}\label{6}\left(\begin{array}{c}
\sum_{m} R_{m1}\\
\sum_{m} R_{m2}\end{array}\right) =\frac{1}{2}
\left(\begin{array}{cc}
1+\sqrt{a} & 1-\sqrt{a}\\
1-\sqrt{a} & 1+\sqrt{a}\end{array}\right) \left(\begin{array}{c}
Q_1\\
Q_2\end{array}\right), \end{equation}
 \begin{equation}\label{7}
\left(\begin{array}{c}
\sum_{n} R_{+n}\\
\sum_{n} R_{-n}\end{array}\right)= \left(\begin{array}{cc}
1 & a\\
0 & 1-a
\end{array}\right)
\left(\begin{array}{c}
P_+\\
P_-
\end{array}\right).\end{equation}
The important feature of (\ref{6}) and (\ref{7}) is that one
marginal contains information only on the standard interference
observable, whereas the other marginal only refers to the standard
path observable. Actually, the bivariate POVM (\ref{16}) was
construed so as to realize this.

Equations (\ref{6}) and (\ref{7}) are applications of a general
definition of a \textit{nonideal} measurement\cite{MadM90}, to the
effect that a POVM $\{M_{i}\}$ is said to represent a nonideal
measurement of POVM $\{N_{j}\}$ if \[
 M_{i} = \sum_{j} \lambda_{ij} N_{j},\;
\lambda_{ij} \geq 0,\;\sum_{i} \lambda_{ij} = 1.\]
 This expression compares the measurement procedures of POVMs
 $\{M_i\}$ and $\{N_j\}$, to the effect that the first can be
 interpreted as a nonideal or inaccurate version of the second,
the nonideality matrix $(\lambda_{ij})$ representing the
nonideality.  A convenient measure of this nonideality is the
average row entropy of the nonideality matrix,
\begin{equation}\label{17}
J_{(\lambda)} := - \frac{1}{N} \sum_{ij} \lambda_{ij} \ln
\frac{\lambda_{ij}} {\sum_{j'} \lambda_{ij'}},\; N \mbox{ \rm the
dimension of matrix }(\lambda_{ij}).\end{equation}

 As is seen from (\ref{6}) and (\ref{7}) the measurement procedure
 depicted in figure~\ref{fig3} gives rise to two nonideality
 matrices, to be denoted by $(\lambda_{mm'})$ and $(\mu_{nn'})$,
 respectively. Under variation of the parameter $a$ the nonideality
 matrices $(\lambda_{mm'})$ and $(\mu_{nn'})$ are seen to
 exhibit a behavior that is very reminiscent of the idea of
complementarity as presented for the first time by Bohr in his
Como lecture\cite{Bohr27}: in one limit ($a=1$) ideal information
is obtained on the standard interference observable, whereas no
information at all is obtained on the standard path observable; in
the other limit ($a=0$) the roles of the standard interference and
path observables are interchanged. For values $0<a<1$ we have
intermediate situations in which information is obtained on the
probability distributions of \textit{both} standard observables,
to the effect that information on one observable gets less
accurate as the information on the other one gets more ideal. By
changing the measurement arrangement so as to also obtain
information on another (incompatible) standard observable, the
information on the first observable gets blurred to a certain
extent.

\subsection{The Martens inequality}\label{sec2.2}
Complementary behavior as discussed above is a rather common
feature of quantum mechanical measurement; many other examples can
be given\cite{dM2002}. Using the nonideality measure
$J_{(\lambda)}$ (\ref{17}) it is possible to give a general
account of this complementarity\cite{MadM90}. Let a bivariate POVM
$(R_{mn})$ satisfy \begin{equation}\label{2.2.1}
\begin{array}{l}
\sum_n R_{mn} = \sum_{m'} \lambda_{mm'} P_{m'},\; \lambda_{mm'}
\geq 0,\; \sum_m \lambda_{mm'} = 1,\\
\sum_m R_{mn} = \sum_{n'} \mu_{nn'} Q_{n'},\; \mu_{nn'} \geq 0,\;
\sum_n \mu_{nn'} = 1,
\end{array}
 \end{equation}
\begin{figure}
\psfig{file=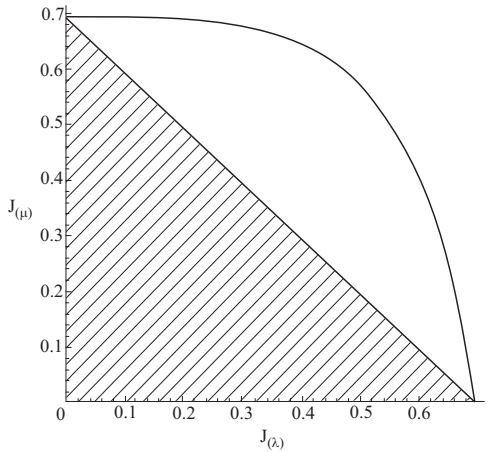,width=2.5in}
 \caption{Parametric plot of $J_{(\lambda)}$
and $J_{(\mu)}$ for the Summhammer, Rauch,Tuppinger experiment.}
 \label{fig4}
\end{figure}
 in which $\{P_m\}$ and $\{Q_n\}$ are maximal PVMs.
Then the corresponding nonideality measures $J_{(\lambda)}$ and
$J_{(\mu)}$ satisfy the \textit{Martens inequality}\footnote{For
nonmaximal PVMs this expression has to be slightly
generalized\cite{MadM90}.}
\[ J_{(\lambda)} +
J_{(\mu)} \geq -\ln \{\max_{mn} Tr P_m Q_n\}.\]
 For the Summhammer, Rauch,Tuppinger experiment
we obtain
\[\begin{array}{l}
J_{(\lambda)}=\frac{1}{2}[(1+a)\ln(1+a)-a\ln a],\\
J_{(\mu)}=\frac{1}{2}[2\ln2-(1+\sqrt{a})\ln(1+\sqrt{a})-
(1-\sqrt{a})\ln(1-\sqrt{a})].
\end{array}\]
In figure~\ref{fig4} a parametric plot is given of these
quantities. The shaded area contains values of $J_{(\lambda)}$ and
$J_{(\mu)}$ forbidden by the Martens inequality. This latter
inequality, hence, does represent the Bohr-Heisenberg idea of
complementarity in the sense of a mutual disturbance of the
information obtained in a joint nonideal measurement of two
standard observables.

\subsection{Martens inequality
versus Heisenberg uncerainty relation}\label{sec2.3} It should be
stressed that the Martens inequality is a general feature of
quantum mechanical \textit{measurement procedures} satisfying
(\ref{2.2.1}). In particular it is independent of the density
operator $\rho$. This feature distinguishes the Martens inequality
from the Heisenberg uncerainty relation
\begin{equation}\label{2.2.2}\Delta A \Delta B \geq \frac{1}{2} | Tr\rho
[A,B]_-|,\end{equation}
 $A$ and $B$ standard observables. It is
interesting to remember that for a long time it has been the
Heisenberg uncerainty relation (\ref{2.2.2}) that was supposed to
describe complementarity in the sense of mutual disturbance in a
joint nonideal measurement of two standard observables. This has
been questioned by Ballentine\cite{Bal70} to the effect that the
Heisenberg inequalities (\ref{2.2.2}) do not refer to
\textit{joint} measurement of incompatible observables at all,
since they can be tested by \textit{separate ideal} measurements
of the two standard observables in question. According to
Ballentine even ``Quantum mechanics has nothing to say about joint
measurement of incompatible observables.''

As far as \textit{standard} quantum mechanics is concerned,
Ballentine is certainly right. However, as is seen from
section~\ref{sec2.2}, the \textit{generalized} quantum mechanics
of POVMs is able to deal with joint nonideal measurement of
incompatible observables. The Martens inequality, rather than the
Heisenberg inequality, is representing the concomitant
complementarity. Although Bohr and Heisenberg had a perfect
intuition as regards the physics going on in a double-slit
experiment, they were not able to give a comprehensive treatment
of it, due to the fact that they did not have at their disposal
the generalized formalism of POVMs. As a consequence they were
restricted to a discussion of the limiting cases only (in our
example $a=0$ and $a=1$). They unjustifiedly thought\cite{dM2000}
that the Heisenberg inequality (\ref{2.2.2}) was the mathematical
expression of their intuition on the intermediate cases (in our
example corresponding to $0<a<1$). However, rather than the
Heisenberg inequality it is the Martens inequality, derived from
the generalized formalism, which serves this purpose. It seems
that, due to a too one-sided preoccupation with measurement, Bohr
and Heisenberg overlooked the possibility that not only
\textit{measurement} but also \textit{preparation} might yield a
contribution to complementarity, the Heisenberg uncertainty
relations referring to the latter contribution.

\section{Bell inequalities}\label{sec3}
As is well known, the Bell inequalities cannot be derived from
\textit{standard} quantum mechanics; they were derived by
Bell\cite{Bell64} from a local hidden-variables theory. As a
consequence, it is generally believed that violation of the Bell
inequalities by the standard Aspect experiments\cite{Asp81,Asp82}
is a consequence of \textit{nonlocality}. In this section it will
be demonstrated that our understanding of the Bell inequalities,
too, can considerably be enhanced by applying the generalized
formalism\cite{dM2002}.

\subsection{Generalized Aspect
experiment}\label{sec3.1} For this purpose the following
experiment is considered, to be referred to as the
\textit{generalized Aspect experiment} (cf. figure~\ref{fig5}).
\begin{figure}
\psfig{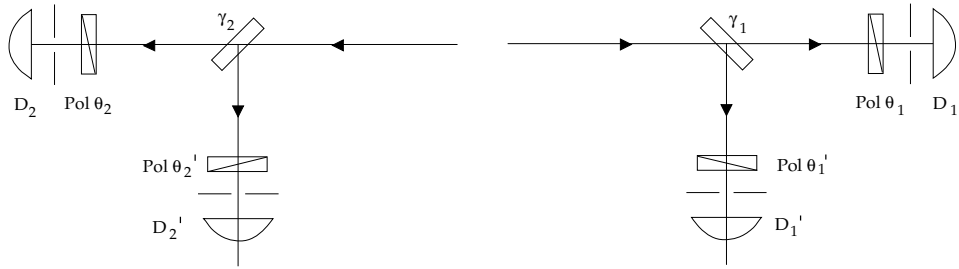}
 \caption{Generalized Aspect experiment.}
  \label{fig5}
\end{figure}
 In the experiment each photon of a
correlated photon pair impinges on a semi-transparent mirror
(transmissivities $\gamma_1$ and $\gamma_2$, respectively). In the
paths of the transmitted and reflected photon beams of photon
$i\;(i=1,2)$ polarization measurements are performed in directions
$\theta_i$ and $\theta'_i$, respectively. Using ideal detectors
$\rm D_1,\;D'_1,\;D_2$ and $\rm D'_2$, a measurement result
(occurrence or nonoccurrence of a click in each of the four
detectors) is obtained for each individual photon pair. The four
(standard) Aspect experiments\cite{Asp81,Asp82} are special cases
of the present experiment, satisfying $(\gamma_1,
\gamma_2)=(1,1)$, $(1,0)$, $(0,1)$ or $(0,0)$, respectively (in
each of these experiments the two detectors which are certain
\textit{not} to click have been omitted).

The generalized Aspect experiment can be analyzed, analogously to
the discussion in section~\ref{sec2}, in terms of complementarity
in the sense of mutual disturbance in a joint nonideal measurement
of incompatible standard observables. Let us first consider the
measurement performed in one arm of the interferometer ($i=1$ or
$2$). In agreement with definition (\ref{2.2.1}) this measurement
can be interpreted as a joint nonideal measurement of the
(standard) polarization observables (PVMs)
$\{{E}^{\theta_i}_+,{E}^{\theta_i}_-\}$ and
$\{{E}^{\theta'_i}_+,{E}^{\theta'_i}_-\}$ in directions $\theta_i$
and $\theta'_i,$ respectively. Thus, by expressing, for a given
$i$, the joint detection probabilities of photon detectors ${\rm
D}_i$ and ${\rm D}'_i$ as $p_{m_in_i}=Tr \rho
R^{\gamma_i}_{m_in_i}$, the bivariate POVM
$(R^{\gamma_i}_{m_in_i})$ is defined according to
\begin{equation}\label{12}(R^{\gamma_i}_{m_in_i}) = \left(
\begin{array}{cc}
O & \gamma_i E^{\theta_i}_+\\
(1-\gamma_i)E^{\theta'_i}_+ &  \gamma_i E^{\theta_i}_- + (1 -
\gamma_i) E^{\theta'_i}_-
\end{array} \right),\;i=1,2.\end{equation}
The marginals of $(R^{\gamma_i}_{m_in_i})$ are found as
 \begin{equation}\label{8}\left(
\begin{array}{c} \sum_{n_i} {R}^{\gamma_i}_{+n_i}\\ \sum_{n_i} {R}^{\gamma_i}_{-n_i}
\end{array} \right)  =
\left( \begin{array}{cc} \gamma_i & 0 \\ 1-\gamma_i  & 1
\end{array} \right)
\left( \begin{array}{c} {E}^{\theta_i}_{+} \\ {E}^{\theta_i}_{-}
\end{array} \right)\; \mbox{\rm (detector D}_i ),\end{equation}
 \begin{equation}\label{9} \left( \begin{array}{c} \sum_{m_i} {R}^{\gamma_i}_{m_i+}\\ \sum_{m_i} {R}^{\gamma_i}_{m_i-}
\end{array} \right) =
\left( \begin{array}{cc} 1-\gamma_i & 0 \\ \gamma_i & 1
\end{array} \right)
\left( \begin{array}{c} {E}^{\theta'_i}_{+} \\ {E}^{\theta'_i}_{-}
\end{array} \right)\;\mbox{\rm (detector D}'_i).\end{equation}
As functions of $\gamma_i,\; 0\leq \gamma_i \leq 1$ the
nonideality matrices in (\ref{8}) and (\ref{9}) show complementary
behavior completely analogous to that of (\ref{6}) and (\ref{7}),
and illustrated by figure~\ref{fig4}. From this complementarity it
can be concluded that, for instance, a measurement result for the
polarization in direction $\theta_i$, obtained in the generalized
Aspect experiment $(0<\gamma_i<1)$, can be different from one
obtained if an \textit{ideal} measurement of this standard
observable $(\gamma_i =1)$ would have been performed. Moreover, it
follows that this difference is a consequence of changing the
measurement arrangement from $\gamma_i=1$ to $\gamma_i <1$, or
vice versa. It is also seen that for $\gamma_i=1$ the marginal
(\ref{9}) is given by $\{O,I\}$, which is a completely
uninformative observable, not yielding any information on the
state of photon $i$. This justifies the omission, referred to
above, of the corresponding detector in the standard Aspect
experiment.

The generalized Aspect experiment depicted in figure~\ref{fig5}
can be interpreted in an analogous way as a joint nonideal
measurement of the \textit{four} standard observables
$\{{E}^{\theta_1}_+,{E}^{\theta_1}_-\},$
$\{{E}^{\theta'_1}_+,{E}^{\theta'_1}_-\}$,
$\{{E}^{\theta_2}_+,{E}^{\theta_2}_-\}$, and
$\{{E}^{\theta'_2}_+,{E}^{\theta'_2}_-\}$, a quadrivariate POVM
being obtained as the direct product of the bivariate POVMs given
in (\ref{12}):
\begin{equation}\label{3.1}{R}^{\gamma_1\gamma_2}_{m_{\!1}n_{\!1}m_2 n_2} =
R^{\gamma_1}_{m_{\!1} n_{\!1}}
 R^{\gamma_2}_{m_2 n_2}.
\end{equation}
From this expression it is evident that there is no disturbing
influence on the marginals in one arm of the interferometer by
changing the measurement arrangement in the other arm. Since
observables referring to different objects do commute with each
other, this is not unexpected. Complementarity in the sense of
mutual disturbance of measurement results is effective in both of
the arms \textit{separately}, disturbance being caused in each arm
by changing the measurement arrangement in that very arm.

\subsection{Complementarity and nonlocality as alternative explanations
of violation of the Bell inequalities}\label{sec3.2}

The interesting outcome of the present discussion of the
generalized Aspect experiment is the existence of a
\textit{quadrivariate} probability distribution
 \begin{equation}\label{11}p^{\gamma_1\gamma_2}_{{m}_{\!1}
{n}_{\!1} {m}_{2} {n}_{2}} = Tr \rho R^{\gamma_1}_{{m}_{\!1}
n_{\!1}}
 {R}^{\gamma_2}_{ {m}_{2} {n}_{2}}
\end{equation}
for the experimentally obtained measurement results. According to
a theorem proven by Fine\cite{Fine}, and by Rastall\cite{Ras83},
the existence of this quadrivariate probability distribution
implies that the Bell inequalities are satisfied by the four
bivariate marginals $p^{\gamma_1\gamma_2}_{{m}_{\!1} {m}_{2}},\;
p^{\gamma_1\gamma_2}_{{m}_{\!1}  {n}_{2}},\;
p^{\gamma_1\gamma_2}_{{n}_{\!1} {m}_{2}}$ and
$p^{\gamma_1\gamma_2}_{{n}_{\!1}  {n}_{2}}$ which can be derived
from (\ref{11}) for fixed $(\gamma_1,\gamma_2)$.

It should be noted that this holds true \textit{also} for each of
the \textit{standard} Aspect experiments, corresponding to one of
the limiting cases $(\gamma_1, \gamma_2)=(1,1),$ etc.. Evidently,
violation of the Bell inequalities by these experiments must be
caused by the fact that no quadrivariate probability distribution
exists from which the four bivariate probabilities
$p^{\gamma_1=1\gamma_2=1}_{{n}_{\!1} {n}_{2}}$,
$p^{\gamma_1=1\gamma_2=0}_{{n}_{\!1} {m}_{2}}$,
$p^{\gamma_1=0\gamma_2=1}_{{m}_{\!1} {n}_{2}}$,
 and $p^{\gamma_1=0\gamma_2=0}_{{m}_{\!1} {m}_{2}}$
 can be derived as marginals. So, the question to be answered is,
 why such a quadrivariate probability distribution does not exist,
 even though for each of the standard Aspect experiments separately
 one is given by (\ref{11}).

A step towards answering this question is the observation that the
quadrivariate probability distributions
$p^{\gamma_1\gamma_2}_{{m}_{\!1} {n}_{\!1} {m}_{2} {n}_{2}}$ given
by (\ref{11}) are different for different $(\gamma_1,\gamma_2)$:
they depend on the measurement arrangement, and so, in general, do
their marginals. Hence, changing the measurement arrangement from
one standard Aspect experiment to another yields a disturbance of
the measurement probabilities, preventing the Bell inequalities
from being derivable from the existence of a single quadrivariate
probability distribution.

In accepting this explanation we may choose between two different
disturbing mechanisms, viz. nonlocality or complementarity. In the
first case it is assumed that the probabilities in one arm of the
interferometer are influenced in a nonlocal way by changing the
measurement arrangement in the \textit{other} arm. This is the
explanation that is generally accepted. The alternative
explanation, based on complementarity, takes into account the
disturbing influence of a change of the measurement arrangement
performed in an arm of the interferometer on the measurement
probabilities measured in that \textit{same} arm, as expressed by
(\ref{8}) and (\ref{9}).

In deciding which of the alternatives, nonlocality or
complementarity, to accept, it is important to realize that, if
four standard observables (PVMs) $\{{E}^{1}_i\}$, $\{{F}^{1}_j\}$,
$\{{E}^{2}_k\}$, and $\{{F}^{2}_\ell\}$ are mutually compatible, a
quadrivariate probability distribution, viz. $Tr \rho
{E}^{1}_i{F}^{1}_j{E}^{2}_k{F}^{2}_\ell$, exists even in the
standard formalism. Hence, \textit{in}compatibility is a necessary
condition for violation of the Bell inequalities. But, since
observables referring to causally disjoint regions of space-time
do commute, only observables referring to the \textit{same} region
can be incompatible. Hence, \textit{in}compatibility is a
\textit{local} affair, as, consequently, is violation of the Bell
inequalities. It seems that complementarity can yield a
\textit{local} explanation of violation of the Bell inequalities,
based on mutual disturbance in the joint nonideal measurements of
incompatible standard observables carried out separately in each
arm of the interferometer. Such an explanation could not be given
on the basis of the standard formalism since, as demonstrated in
section~\ref{sec2}, that formalism is not able to deal with this
kind of complementarity. Dependence on the measurement arrangement
is only evident when considering the bivariate probabilities $Tr
\rho R^{\gamma_i}_{m_in_i}$, derived from (\ref{12}), which do not
exist in the standard formalism.

\bibliographystyle{ws-procs975x65}
\bibliography{procleiden}

\end{document}